\documentstyle[aps]{revtex}
\draft

\begin{document}

\title{Comment on "Bose-Einstein condensation in
low-dimensional traps" }
\author{Wu-Sheng Dai \thanks{daiwusheng@tju.edu.cn}}
\address{Department of Applied Physics, Tianjin University, Tianjin
300072, P. R. China}
\author{Mi Xie \thanks{xiemi@mail.tjnu.edu.cn}}
\address{Department of Physics, Tianjin Normal University,
Tianjin 300074, P. R. China}
\date{}
\maketitle

\begin{abstract}

We show that the critical temperature of a one-dimensional gas confined by a
power-law potential should be lower than that in the paper of Vanderlei
Bagnato and Daniel Kleppner. Moreover, a sketch of the critical temperature
is given in some more details.
\end{abstract}

\vskip 1cm

The paper by Vanderlei Bagnato and Daniel Kleppner discusses the
Bose-Einstein condensation of an ideal Bose gas confined by one- and two-
dimensional traps \cite{Bagnato}. This is an important work and the 
results are widely cited \cite{griffin}.

The density of states of a one-dimensional gas confined by a potential $%
U(x)=U_{0}(\left| x\right| /L)^{\eta }$ is

\begin{eqnarray*}
\rho \left( \varepsilon \right) &=&\frac{\sqrt{2M}}{h}\int_{-l(\varepsilon
)}^{l(\varepsilon )}\frac{dx}{\sqrt{\varepsilon -U(x)}} \\
&=&\frac{2}{\eta }\frac{\sqrt{2M}}{h}L\frac{\varepsilon ^{\frac{1}{\eta }-%
\frac{1}{2}}}{U_{0}^{\frac{1}{\eta }}}F(\eta ),
\end{eqnarray*}%
where $l(\varepsilon )=L(\frac{\varepsilon }{U_{0}})^{\frac{1}{\eta }}$, and

\[
F(\eta )=\int_{0}^{1}\frac{y^{\frac{1-\eta }{\eta }}dx}{\sqrt{1-y}}=\frac{%
\sqrt{\pi }\Gamma (\frac{1}{\eta })}{\Gamma (\frac{1}{\eta }-\frac{1}{2})}. 
\]
In ref. \cite{Bagnato} the factor $\frac{2}{\eta }$ is lost.

The number of particles in the excited states is given by

\begin{eqnarray*}
N_{e} &=&\int_{0}^{\infty }d\varepsilon \rho \left( \varepsilon \right) 
\frac{1}{e^{(\varepsilon -\mu )/kT}-1} \\
&=&\frac{2}{\eta }\frac{\sqrt{2M}}{h}L\frac{F(\eta )}{U_{0}^{\frac{1}{\eta }}%
}(kT)^{\frac{1}{\eta }+\frac{1}{2}}g_{1}(\eta ,\frac{\mu }{kT}),
\end{eqnarray*}
where

\vspace{1pt} 
\[
g_{1}(\eta ,x)=\int_{0}^{\infty }dy\frac{y^{\frac{1}{\eta }-\frac{1}{2}}}{%
e^{y-x}-1}. 
\]

\vspace{1pt}The Bose-Einstein condensation begins when $N_{e}$ equals to the
total number of particles in the system $N$. In this case the chemical
potential $\mu =0$. The critical temperature $T_{c}^{1D}$ can be given by
the expression:

\[
kT_{c}^{1D}=\left( \frac{\eta }{2}\right) ^{\frac{2\eta }{2+\eta }}\left[ 
\frac{N}{L}\frac{h}{\sqrt{2M}}\frac{U_{0}^{\frac{1}{\eta }}}{F(\eta )}\frac{1%
}{g_{1}(\eta ,0)}\right] ^{\frac{2\eta }{2+\eta }}. 
\]%
As a result of adding $\frac{2}{\eta }$ in the density of states $\rho
\left( \varepsilon \right) $, a factor $\left( \frac{\eta }{2}\right) ^{%
\frac{2\eta }{2+\eta }}$ appears in the expression of the critical
temperature $T_{c}^{1D}$. According to the discussion in \cite{Bagnato}, 
the one-dimensional gas displays Bose-Einstein condensation only when the
potential power $\eta <2$. The value of the factor $\left( \frac{\eta }{2}%
\right) ^{\frac{2\eta }{2+\eta }}$ is always less than $1$ when $0<\eta <2$.
Therefore, the critical temperature will be suppressed by such a factor. The
variation of $T_{c}^{1D}$ with $\eta $ is shown in Fig1. In ref. \cite{Bagnato}, 
the curve of the critical temperature $T_{c}^{1D}$ does not
include the case of small $\eta $. This makes $T_{c}^{1D}$ looks like a
monotonically decreasing function of $\eta $. In Fig1, the dependence of $%
T_{c}^{1D}$ on $\eta $ is drawn in the whole interval $[0,2]$. The critical
temperature is found to show a peak between $\eta =0$ and $2$.

\vskip 2cm

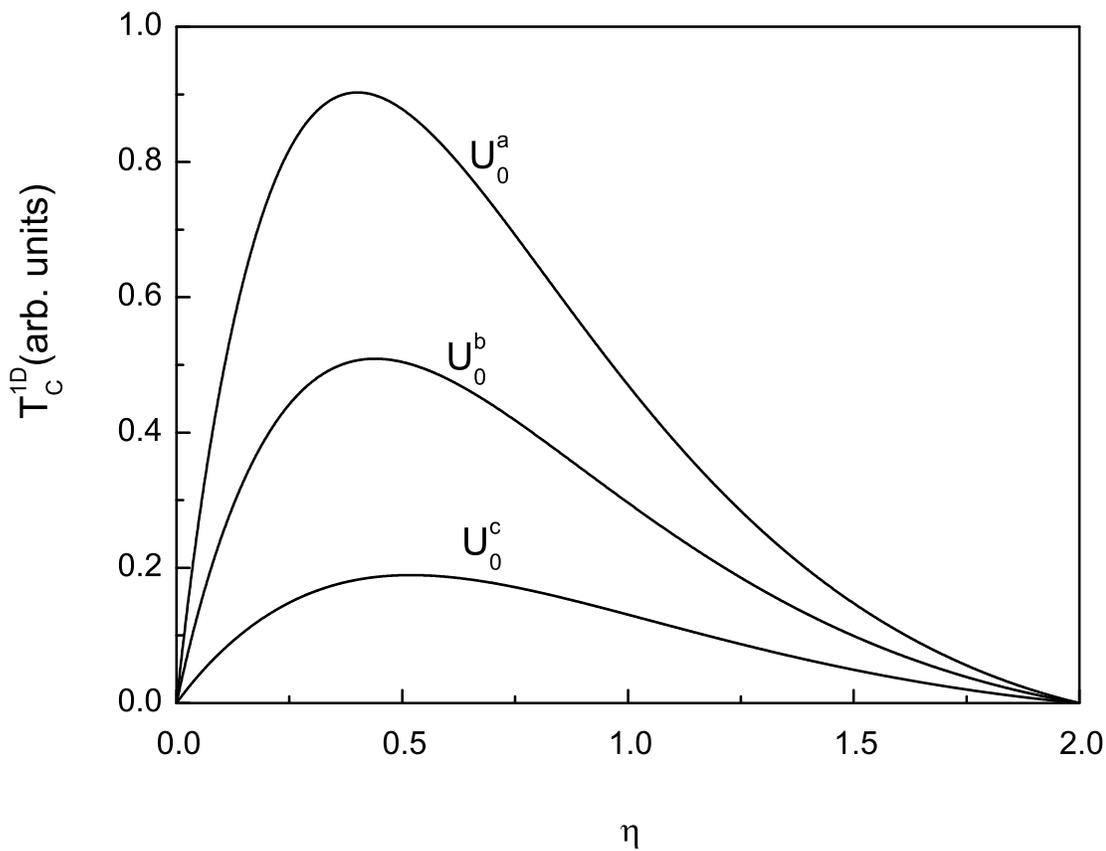
Fig1. Critical temperature $T_{c}^{1D}$, as a function of the potential
power $\eta $, for various values of $U_{0}$ ($U_{0}^{a}>U_{0}^{b}>U_{0}^{c}$%
).
\end{document}